**Revisiting in-depth chemical analysis for alkali ions-exchanged glass**

Se-Ho Kim[a,*], Leigh T. Stephenson[a], Torsten Schwarz[a], Baptiste Gault[a,b*]

[a] Max-Planck-Institut für Eisenforschung, Max-Planck-Straße 1, 40237 Düsseldorf, Germany

[b] Department of Materials, Royal School of Mines, Imperial College London, London, SW7 2AZ, United Kingdom

*co-corresponding authors: s.kim@mpie.de, b.gault@mpie.de

**Abstract**

The upcoming flexible ultra-thin glass for foldable displays has attracted widespread attention as an alternative to rigid electronic smartphones. However, the detailed compositional effect of the chemical strengthening of the glass is not well understood. The spatially resolved chemistry and the depth of compression layer of tempered glass is far from being clear. In this study, commonly-used X-ray spectroscopy techniques and atom probe tomography (APT) were used comparatively to investigate the distribution of constituent elements in two representative smartphone glass samples: non-tempered and chemically-tempered glasses. Using APT, analysing the alkalis (Li, Na, K, Ca) that could impact the failure of the tempered glass was feasible at sub-nanoscale resolution; demonstrating that APT can be regarded as an alternative technique for imaging chemical distribution in glass for mobile applications.

## Introduction

The chemical strengthening of glass was initially proposed by Kistler in 1962 (KISTLER, 1962). This process involves replacing one of the constituents by an element with a different atomic size and electronic polarizability, while maintaining the structure of the pristine glass. Since then, extensive research and industrial efforts have been carried out to develop this approach for mechanical strengthening of glasses for a wide range of applications including cockpit windows for aircrafts, digital dashboards and passenger seat displays in automobiles, hard disk drives for data storage (Gy, 2008). Most importantly, it is seen as the best fabrication method to strengthen ultra-thin glass (UTG) sheets efficiently without influencing their optical properties and mass (Yuan et al., 2021). UTG, with a thickness below 100 μm, is a rising display material that can be folded or rolled like a flexible polymer, while providing protection of the screen (Tanaka, 2013). Promises for UTG application range from wearable devices to solar cells and even flexible smartphone (SCHOTT, 2021). A recent commercial flip phone already adopted UTG glass with a thickness down to 30 μm on its rigid display (SamsungDisplay, 2020) and UTGs with a thickness of ~5 μm has demonstrated good flexibility in micro-scale applications such as a femtoliter nanofludic valve (Kazoe et al., 2019) and a diaphragm pressure transducer (Yalikun & Tanaka, 2017).

The replacement of small ions (e.g., $Na^+$) by larger ones from an external ion source (e.g., $K^+$) creates a compressive stress at the surface that improves the mechanical strength, preventing the possible formation of micro/nano-cracks on the glass surface (NORDBERG et al., 1964). However, if the ion exchange process continues for a bulky glass, the surface compressive stress will increase while the tensile stress inside the glass will rapidly increase and a complete diffusion throughout the thickness could create severe internal-compressive stress resulting in failure of the material (Terakado et al., 2020; Shan et al., 2018). Therefore, the depth of ion-diffused layer (DOL), defined as the depth at which the residual stress is equal to zero, must be measured in order to obtain high quality of the chemically strengthened glass. As a rule of thumb, the DOL limitation for UTG is generally thought to be one-sixth of the glass thickness, which is calculated to be 830 nm as the threshold depth for a 5-μm-thickness UTG.

Despite the fact that the kinetics of the alkali ion interdiffusion process and chemical strengthening are well reported, a better understanding of the underlying atomic-scale compositional effect is unclear, especially for UTG (Varshneya, 2010). This is due to the limitation of in-depth chemistry characterization instruments offering site-specific analysis. As a result, most research still relies on spectroscopy techniques or the nominal weight compositions from the powder-melting synthesis (Kim et al., 2021; Wang et al., 2014; Farah et al., 2014). Here, we assess the performance of atom probe tomography (APT)(Gault et al., 2021) in a systematic comparison with more conventional chemistry-analysis tools for chemical analysis of an aluminosilicate tempered glass, and correlate it with the increase in the glass hardness. We discuss the strength and weaknesses of these selected techniques and demonstrate the complementarity offered by APT that provides spatially-resolved alkali-(earth) elements' distributions at the sub-nanometre level, including for light and trace elements.

## Materials and Methods

*Materials*

We chose an aluminosilicate glass with an initial thickness of 750 μm that is used for a mobile display. A first glass was without any further chemical or heat treatment and is labelled as 'Glass_0'. A second sample was subjected to ion-exchange treatment by using an industrial-scale furnace that allows the treatment of series of glasses and is named as 'Glass_1'.

*Ion exchange*

In the ion-exchange process, the submerging time of the pristine sample into the alkaline solution is calculated based on the diffusion law. Usually, a batch temperature of below 460 °C and a processing duration below 24 hrs are recommended (Bartholomew & Garfinkel, 1980). A potassium nitrate ($KNO_3$) salt bath was prepared by heating an industrial source (>99.9%) of $KNO_3$ up to around 460 °C. The sample was submerged into bath in the furnace for an hour to achieve an equilibrium state (sample labelled 'Glass_1').

Taking into account a constant diffusivity of $K^+$ in aluminosilicate glass, the $K^+$ concentration along the sample is constant as long as the alkaline solution is not depleted (acting as a reservoir). Using planar doping of the $K^+$ ion, the concentration profiles can be derived from a one-dimensional diffusion solution corresponding to Fick's second law equation: $C_{K^+} = C_{K^+,0}\left(1 - \mathrm{erf}\left(\frac{l}{2\sqrt{D_{K^+}t}}\right)\right)$, where $C_{K^+}$ and $C_{K^+,0}$ are surface concentration of K ion at time, t, and in the initial state, respectively. $D_{K^+}$ is the concentration-independent diffusivity parameter of $K^+$ in aluminosilicate glass that is reported to be $1.5 \times 10^{-11}$ cm$^2$ sec$^{-1}$ at 460 °C (Karlsson et al., 2017, 2010). To reach the equilibrium diffusion state, $C_{K^+}/C_{K^+,0}$ must be near one, which gives the error function variable $\frac{l}{2\sqrt{D_{K^+}t}} \cong 3$. It is reported that the optimal depth of an ion-exchanged layer for a mobile display window (~500 μm) is between 12 to 22 μm; therefore, we chose 20 μm (Ahn et al., 2014). The tempering time, t, value is calculated to be 2 hrs.

*Mechanical testing*

Two 2x2 cm$^2$ samples were prepared from 750 μm thickness blank samples: Glass_0 and _1 glasses. Nano-indentations tests were done using a Berkovich indenter with an apex semi-angle of 65° and with a constant indenter load of 9 mN. 100 indentations were made on each specimen and a distance of 5 μm was kept between the center of each indentation.

*Characterization*

The Glass_0 and _1 glasses samples were characterized with respect to their composition, structure and chemical bonding vibrations using inductively-coupled-plasma optical-emission spectroscopy (ICP-OES, Agilent 5800) and Raman spectroscopy (WiTec: Alpha300) in the range of 200 to 2000 cm$^{-1}$ with a green laser source (wavelength: 532 nm).

Scanning electron microscope (SEM)-energy dispersive X-ray spectroscopy (EDS) elemental analysis (FEI Helios Nanolab 600i) was performed on the Au-coated glass samples at 30 kV accelerating voltage and 2.4 nA current.

The surface region was prepared into a series of APT specimens by using a site-specific preparation approach described in Ref. (Thompson et al., 2007) using a Ga-focused ion beam (FIB) / SEM (FEI Helios Nanolab 600i) (see Fig. 1). APT analyses were performed on both samples by using local electrode atom probes (CAMECA reflectron-fitted LEAP 5000 (XR) and straight-flight-path LEAP 5000 (XS) systems) in pulsed UV laser mode at a detection rate of 1 %, a laser pulse energy of 90 pJ, and a pulse frequency of 125 kHz. The specimen temperature was set to 60 K during analysis. Data reconstruction and analyses were performed using the commercial software Imago visualization and analysis system standard (IVAS) 3.8.2 developed by CAMECA Instruments. Note that a low composition below 0.5 at.% Ga stemming from the focused-ion beam preparation was detected but when Xe FIB system (Helios plasma-FIB Thermo-Fisher) was used for the APT specimen preparation negligible amount of Xe was detected in the glass specimen.

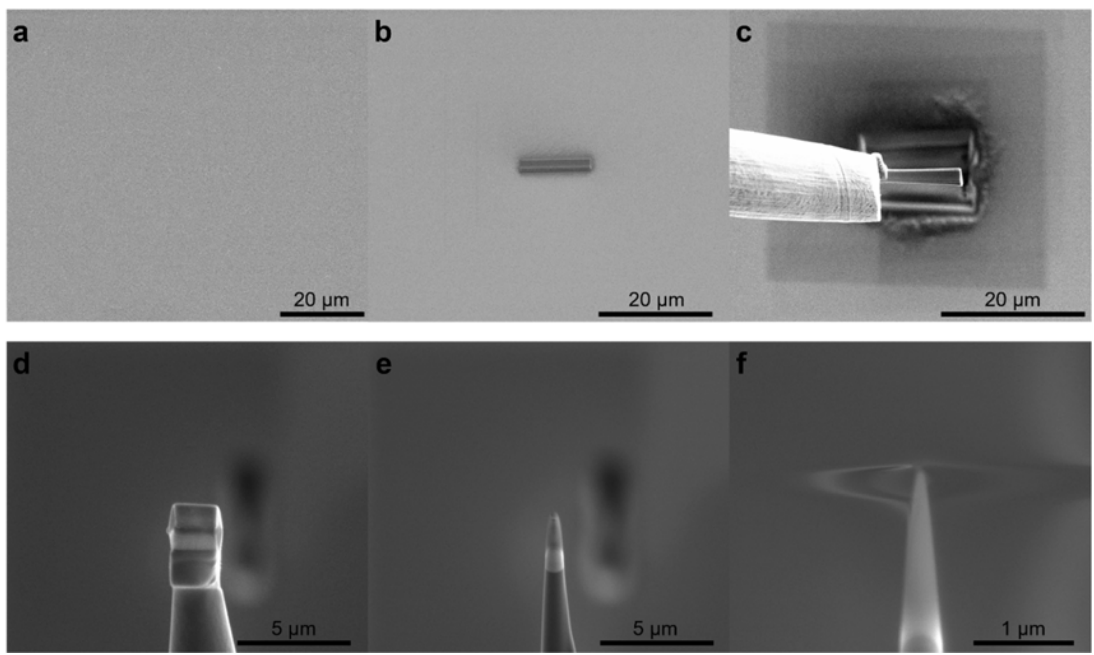

**Fig. 1.** APT specimen preparation: (a) surface of the Glass_0 sample, (b) FIB Pt-C coating, (c) the lift-out process using an omniprobe, (d) welding on a Si micro-post, (e) Annular milling process, (f) Final needle-like APT specimen.

**Results and Discussion**

*Mechanical testing*

The Glass_0 and the 2hrs-chemical-treated glass, Glass_1, are displayed in Fig. 2a. Nano-indentations was performed on both samples and as expected, the hardness of Glass_1 showed a 7.3% higher value than that of Glass_0 (Fig. 2b). Raman spectroscopy is often used to identify the chemical structure in tempered glass, but in our case, despite a sufficient amount of time for a complete ion-exchange process, no clear indication of peak shifts nor intensification between samples was measured (Fig. 2c). The broad Raman peak at 1050 cm$^{-1}$ corresponds to the mixed signal of the stretching $T_2$ vibrational mode of Al/Si-$O_4$ tetrahedra and stretching Al/Si-O of the different Q-bands units (Le Losq et al., 2014; Malfait et al., 2008; Mysen, 1990). $D_1$ (at 480 cm$^{-1}$) and $D_2$ (at 580 cm$^{-1}$) peaks correspond to symmetric stretching of O atoms in four and three-membered tetrahedral rings, respectively (Terakado et al., 2020; Le Losq et al., 2014).

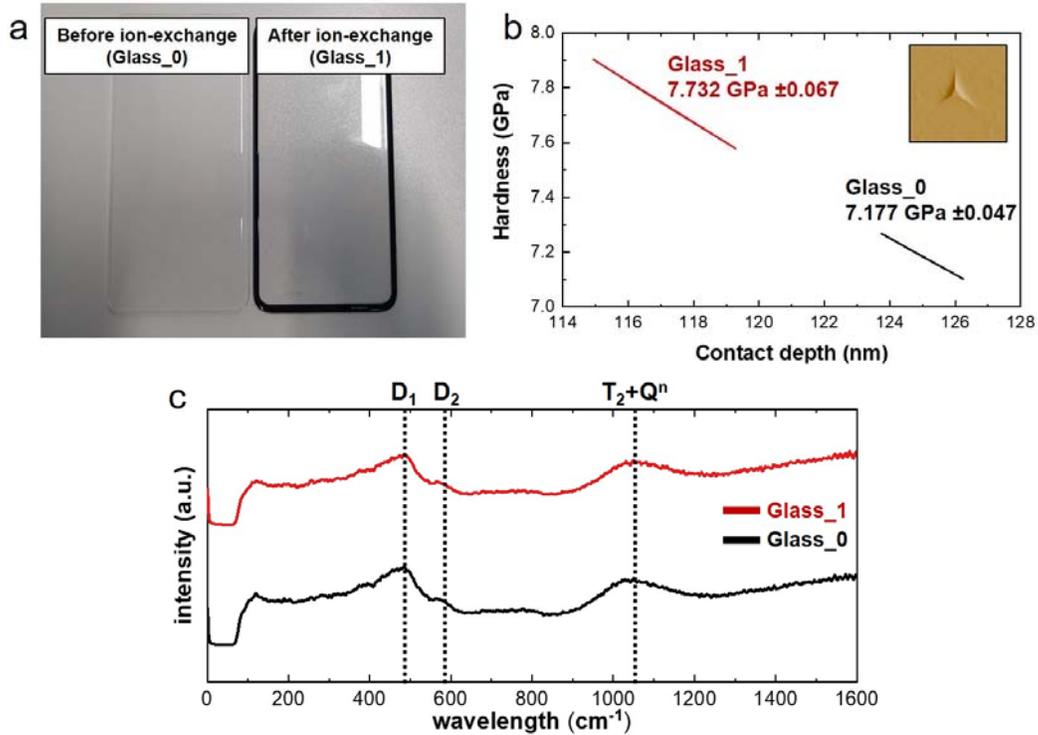

**Fig. 2.** (a) As-received glass and chemical tempered glass. (b) Vickers hardness (LECO M-400) tests. (c) Raman spectroscopy (WiTec: Alpha300R) results.

*Bulk characterization*

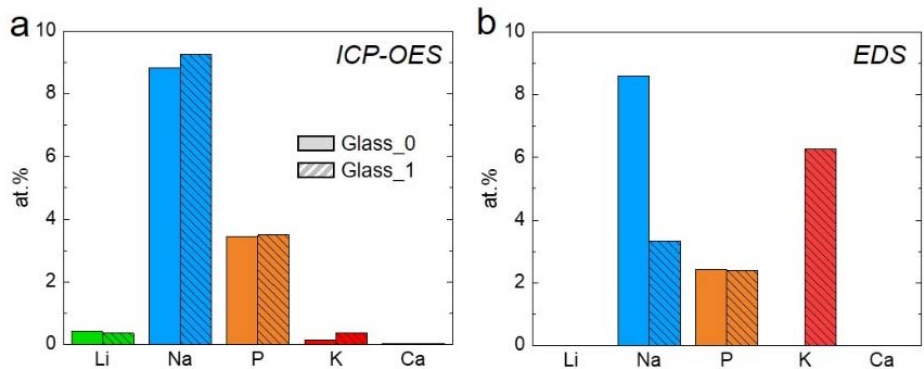

**Fig. 3.** Composition comparison plots of Li, Na, P, K, and Ca for Glass_0 and Glass_1 from (a) ICP-OES and (b) SEM-EDS.

ICP-OES is commonly used for chemical analysis of tempered glass (Calahoo et al., 2016; Schenk & Almirall, 2012). The dissolved liquid glass is heated with plasma emitting light which is then measured with the spectrometry. After the exchange process, an increase of K content is observed but the overestimation of the Na content after the exchange disagrees with the expected result (Fig 3a and Table 1). A possible explanation could be the introduction of Na during the preparation of the specimens. Each glass display was crushed into a powder with a mortar and pestle, followed by dissolution with hydrofluoric acid. Moreover, the mortar and pestle constituent element (*e.g.* Zr) was also detected implying possible introduction of impurities during the sample preparation, yet this was not detected from other analytical techniques.

**Table 1.** Atomic composition of elements for Glass_0 and _1 measured by ICP-OES, SEM-EDS, and APT-HR, XS. The oxygen content for ICP-OES was calculated by normalizing from the detected composition.

| at.% | O | Si | Al | Li | Na | P | K | Ca | impurities from sample preparation | | | |
|---|---|---|---|---|---|---|---|---|---|---|---|---|
| | | | | | | | | | Ga | Au | Zr | C |
| Glass_0 | | | | | | | | | | | | |
| ICP-OES | (36.96) | 34.66 | 15.26 | 0.42 | 8.82 | 3.46 | 0.13 | 0.032 | | | 0.25 | 0.01 |
| SEM-EDS | 34.14 | 34.95 | 16.68 | - | 8.59 | 2.43 | - | - | | 3.21 | | |
| APT_XR | 56.47 | 21.95 | 11.52 | 4.87 | 3.25 | 1.85 | 0.017 | 0.023 | 0.048 | | | |
| APT_XS | 55.61 | 20.81 | 12.72 | 4.47 | 4.43 | 1.46 | 0.015 | 0.027 | 0.47 | | | |
| Glass_1 | | | | | | | | | | | | |
| ICP-OES | (36.95) | 34.36 | 14.99 | 0.37 | 9.26 | 3.50 | 0.37 | 0.028 | | | 0.23 | 0.01 |
| SEM-EDS | 42.07 | 30.35 | 14.42 | - | 3.34 | 2.40 | 6.26 | - | | 1.15 | | |
| APT_XR | 57.15 | 22.86 | 12.31 | 1.97 | 1.96 | 1.76 | 1.69 | 0.050 | 0.24 | | | |
| APT_XS | 55.95 | 21.53 | 14.03 | 2.68 | 2.10 | 1.37 | 2.24 | 0.056 | 0.054 | | | |

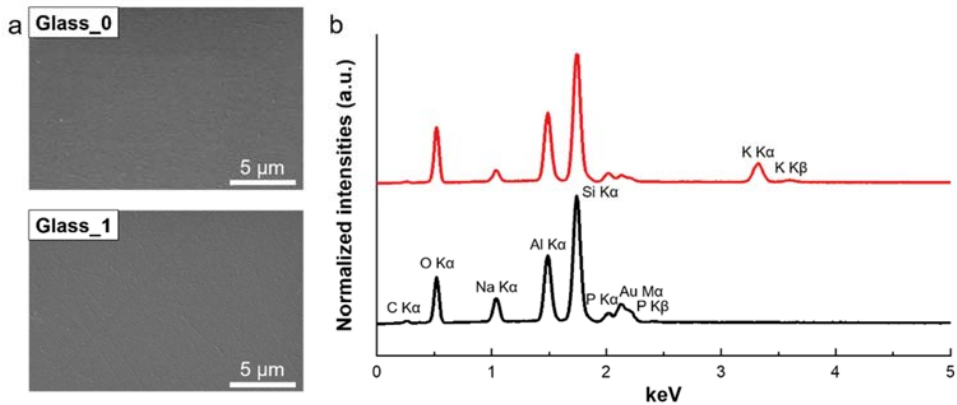

**Fig. 4.** (a) SEM surface images of Au-coated Glass_0 and _1 samples. (b) Corresponding SEM-EDS spectra. A thin layer of Au was coated prior the EDS-SEM measurements thus the Au signal was not from the sample.

SEM-EDS elemental analysis is also commonly used for chemical composition analysis of glasses (Fourmentin et al., 2021; Erdem et al., 2017). With the SEM-EDS, ~2 μm depth resolution is possible, and is typically sufficient for near-surface chemistry analysis (EDAX, 2020). Fig. 4 shows the surface SEM images and EDS spectra of the Glass_0 and _1. The atomic fraction of the detected trace elements (Na, P, and K) is plotted in Fig. 3b. The decrease in Na after the ion exchange process indicates the at Na was replaced by K in the analysed region. However, no information regarding other elements (Li and Ca) was obtained due to the elemental detection limit for trace amount (<0.5 wt%), in particular for light elements (Leng, 2013). This representative result may not provide sufficient details to advance understanding of the chemical exchange effect.

*Atom probe tomography*

We performed APT to evaluate the composition and distributions of elements in both samples. Unlike X-ray-based spectroscopy or other bulk mass spectrometry techniques, APT does not require corrections by relative sensitivity factors that can vary by orders of magnitude depending on the considered element, and, despite known issues in the detection of e.g. oxygen (Zanuttini et al., 2017; Gault et al., 2016; Saxey, 2011) and alkalis (Lu et al., 2017; Greiwe et al., 2014) in oxides, APT has the potential to provide local analysis with his chemical sensitivity in combination with sub-nanometer spatial resolution for detecting local deviation in composition (Wang et al., 2016; Gin et al., 2013; Lu et al., 2017).

Typically, different acquisition parameters (e.g. base temperature, laser-pulse energy) impact the APT performance, including the background level and eventually the mass and spatial resolutions. Especially in the case of poorly- or non-conductive materials (*e.g.* glass), the measurement parameters require careful optimization (see Fig. 5). The laser parameters were swept every 2.5 million ions for a glass specimen. In general, pulsing at low laser frequency requires time to collect a sufficient number of ions to ensure statistically relevant analysis and evaporating ions at high laser power develops thermal tails. So, 125 kHz and 90 pJ were used for the optimal parameter conditions on the LEAP 5000XR system.

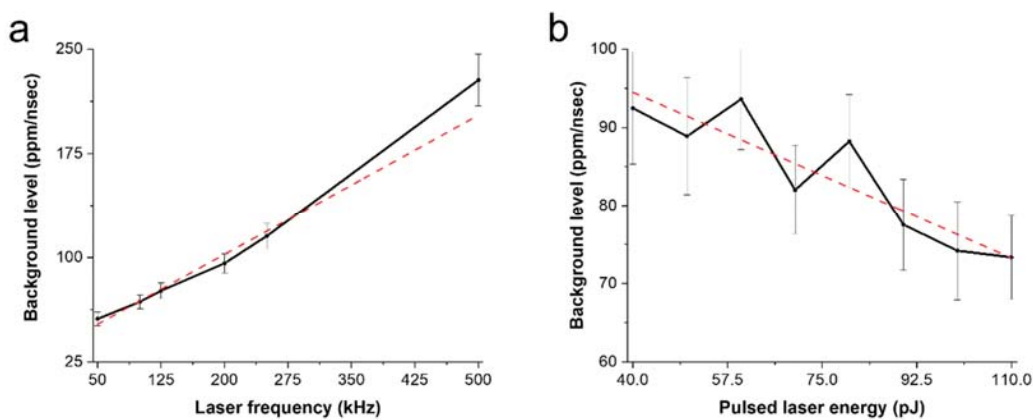

**Fig. 5.** Background levels *vs.* (a) laser frequency and (b) pulsed-laser energy.

Examples of mass spectra obtained from both instruments are shown in Fig. 6a and 6b, respectively. The former has a detection efficiency of approx. 80% whereas the latter, only approx. 50%. In contrast, the XR system has a higher mass resolution, as the reflectron enables longer flight times leading to better separation of mass peaks. For instance, in the XR system, a clear peak at 15.5 Da corresponding to $^{31}P^{2+}$ ion was detected (insets in Fig. 6) whereas the XS system could not resolve it as readily. No significant difference in the overall atomic composition was found between two instruments (Table 1) but we used the HR system for detailed chemical analysis in the acquired mass spectrum.

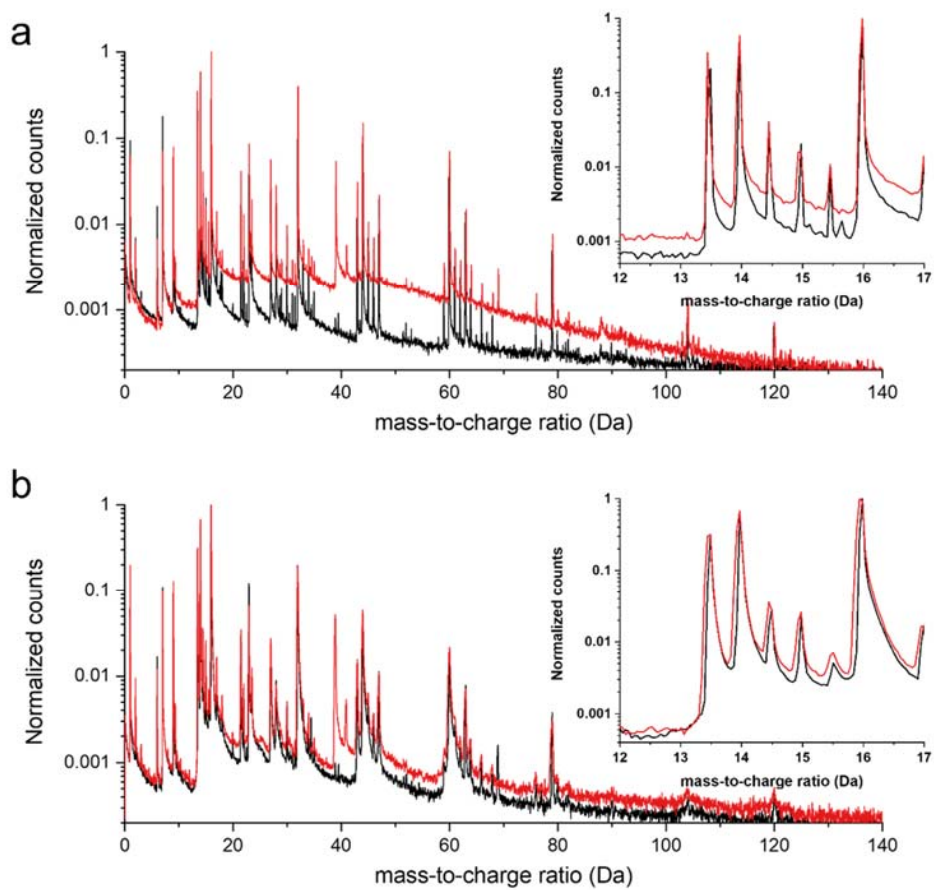

**Fig. 6.** Mass spectra of (black) Glass_0 and (red) Glass_1 samples from (a) 5000 HR and (b) 5000 XS atom probe instruments. Insets show the extracted mass spectrum regions at 12-17 Da on each dataset.

As mentioned earlier, quantification by APT for ceramic sample a is challenging. The dataset from Glass_2 in the LEAP 5000 XS was further analysed to investigate the multiplicity of the detected ions, in order to see if additional chemical information could be retrieved from the otherwise un-indentified ions. In Fig. 7a, the correlation histogram of the multiple-ion events shows that there was continuous field evaporation of complex oxygen-containing molecular ions caused by the electrostatic field, i.e. not triggered by the temperature increase associated to the laser pulse that can be considered as DC evaporation, or with a very long delay following this thermal pulse. These ions are the main contributors to the high level of background. Despite strong and clearly discernible peaks throughout the mass-to-charge spectrum of the tempered state, the background still holds a large remaining portion (30%). The single (M=1) spectrum has a slightly different distribution to the multiple (M>1) spectrum (see Fig. 7b). This difference could be attributed to a distinct component corresponding to a higher fraction of DC evaporation events. This has already been discussed previously by Yao et al. for instance(Yao et al., 2010, 2013). Nevertheless, our interest is primarily in the detection of alkali ion species that do not seem to be affected much compared to Al, Si, and mostly O ions. This may be related to their high electropositivity that make them less prone to forming molecular ions under intense positive electric field.

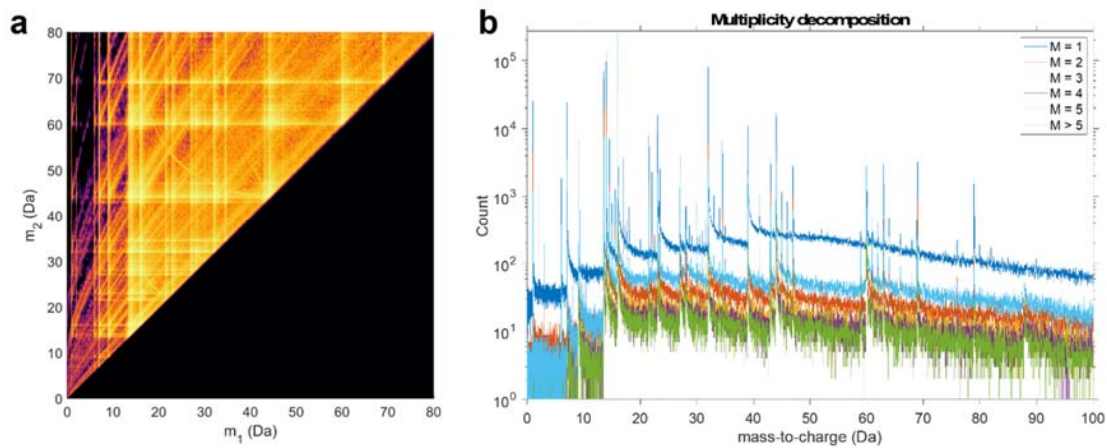

**Fig. 7.** An issue in the mass spectroscopy of oxide materials arises from evaporation uncorrelated with the laser pulse. (a) Correlation histogram of the multiple hits. (b) Multiplicity decomposition of single, double, triple, quadruple, and quintuple hits.

Fig. 8 shows the overall reconstructed 3D atom maps of Glass_0 (up) and Glass_1 (down). Major elements (Al, Si, O), minor elements (P, Li, Na, K), and trace element of Ca (not shown in the atom maps) are detected. All elements are homogeneously distributed and no enrichment nor depletion are observed. To investigate the chemistry and structure further, the inner regions of the atom maps are extracted as shown in Fig. 9a: Glass_0 (right) and Glass_1 (left).

After the exchange, alkali elements such as Li, Na, and K are detected in the Glass_1. The distributions of the alkali ions are assessed by using a nearest-neighbour analysis, and they appear comparable to a random distribution (Shariq et al., 2012; Stephenson et al., 2007) (Fig. 9b and 9c), $Li^+$ and $Na^+$ levels decreased by 2.5 and 1.6 times, respectively, whereas the content of $K^+$ increased up to 1.7 at.%. Interestingly the drop in $Li^+$ concentration is larger than for $Na^+$. This can be simply explained by the higher diffusivity respect to ion radii: $Li^+$ has the smallest size of 78 pm whereas $Na^+$ and $K^+$ have the size of 98 and 133 pm, respectively. The $Li^+$– $K^+$ ions replacement will have a greater size mismatch and thus a greater compressive stress is formed when compared to $Na^+$–$K^+$ exchange. For Li-doped borosilicate, Li electromigration was observed and Li ions continuously move towards the specimen apex during the field evaporation process (Greiwe et al., 2014). In contrast, here in the aluminosilicate glass, despite long measurements (50 M ions), no migration of Li or other elements was observed (see Fig. 8 and 9c). The atomic ratio of Si to Al for both glasses is ~2, which indicates that the molecular ratio of $Al_2O_3$ to $SiO_2$ compounds is 1:1. This agrees with the one-to-one ratio of $Al_2O_3$ and $SiO_2$ was also measured with other analytical instruments, as summarised in Fig 10a.

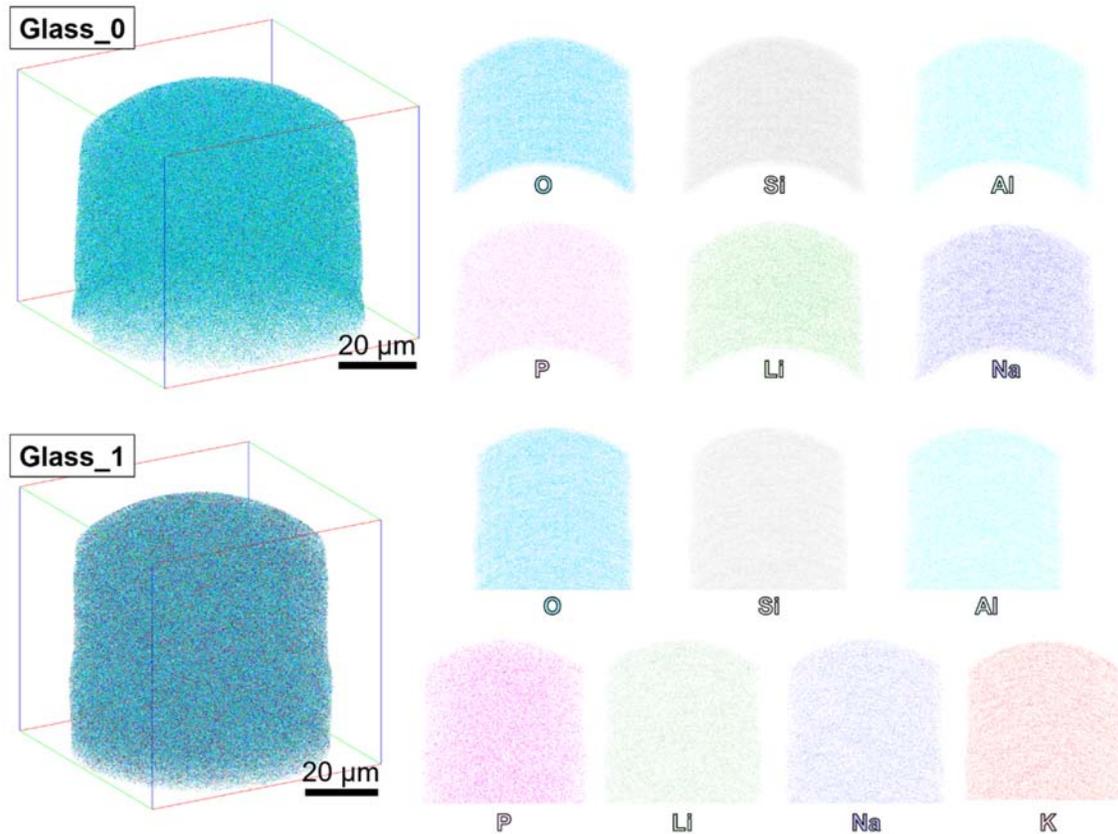

**Fig. 8.** 3D atom maps of (top) Glass_0 and (bottom) _1 samples. The APT measurements on semiconductor and ceramic materials show a consistent deficiency in O which cannot able to be removed by parameter optimization. It is postulated that the fragmentation of evaporation molecular oxide ions may play a role in producing neutral O atoms that cannot cascade the signals on the analysis detector. Nevertheless, in this work, the APT results show high accuracy of O content comparing with other analytical measurements.

Besides, P and Ca are also detected, with 1.85 at.% and 230 appm, respectively in Glass_1. $PO_x$ in aluminosilicate is known to increase compressive stresses and its increase in concentration effectively promotes $Na^+$ and $K^+$ inter-diffusivity (Zeng et al., 2016; Mysen, 1998). Nevertheless, it is susceptible to water or moisture and can be leached even at room temperature (Tošić et al., 2013). The P contents in Glass_0 and _1 are 1.85 and 1.76 at.% and a slight decrease of P by 5% in the Glass_1 is observed which can be due to the washing process in water after the chemical tempering process in molten salt.

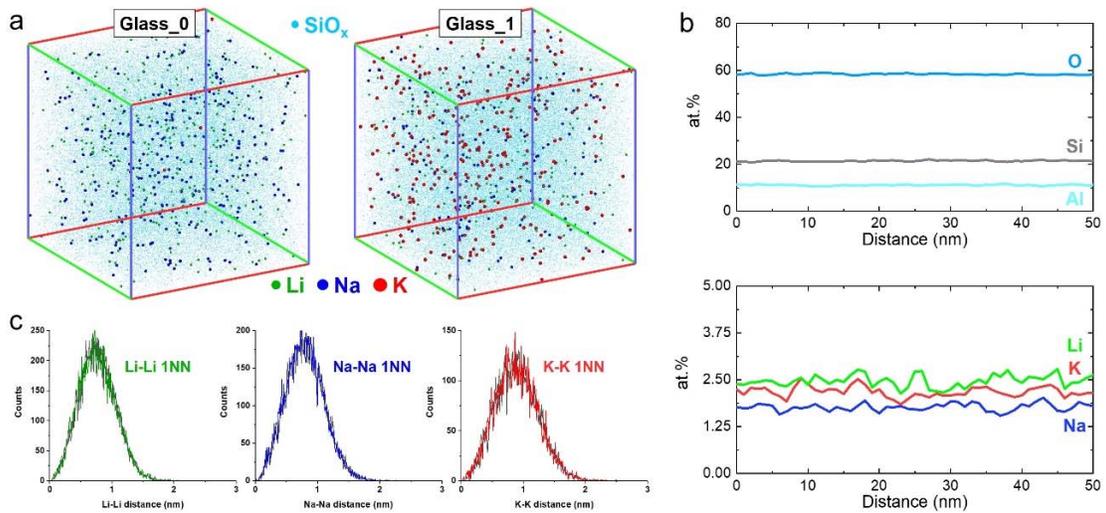

**Fig 9.** (a) 3D reconstructed atom maps (35x35x35 nm$^3$) of the Glass_0 and _1. Light blue, green, blue, and red dots represent SiO$_x$ complexes, Li, Na, and K. P and Ca atoms are not shown. Each overall dataset is presented in Fig. 7. (b) 1D atomic compositional profile of the Glass_1 along the measurement direction ($\phi$15nm x 50 nm$^3$ and bin width of 1 nm). (c) The experimental and random (grey) alkali-alkali elements nearest-neighbour (1NN) distribution.

*General discussion*

In the chemical-tempered glass industry, the presence of Ca cations in concentrations as low as 10 ppm in the ion-exchange KNO$_3$ bath can cause major problems (Sglavo, 2015). Ca$^{2+}$ penetrates into the silicate compound surface, inhibits the Na$^+$-K$^+$ exchange, and consequently, prevent the generation of the necessary compressive stress (Sglavo et al., 2017; Xiangchen et al., 1986). Ca has an ionic radius (100 pm) very similar to Na$^+$ and smaller than K$^+$; therefore, replacing with Na$^+$ is thermodynamically favourable on the glass surface. Although a high purity level of molten KNO$_3$ was used, the Ca content in the glass after the chemical tempering was increased by a factor of two. The comparison of minor elements compositions in between before and after chemically-tempered glasses is summarized in Fig. 10b.

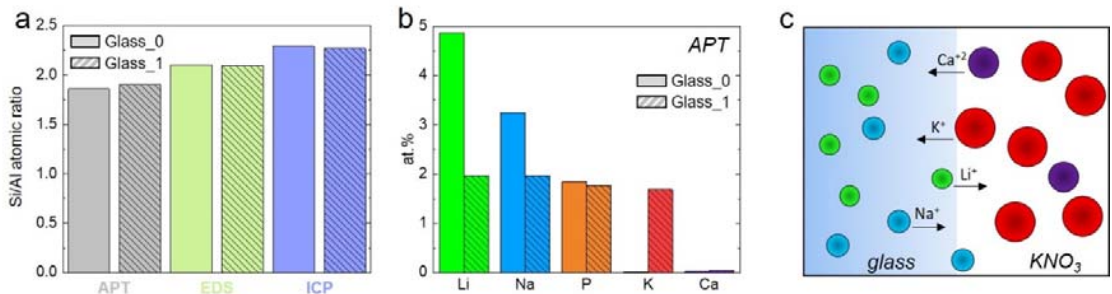

**Fig. 10.** (a) Si/Al ratio of Glass_0 and _1 from different analytical instruments: APT, SEM-EDS, ICP-OES. Composition comparison plots of Li, Na, P, K, and Ca for Glass_0 and Glass_1 from APT. (c) Schematic illustration of alkali(earth) ions diffusion during the glass tempering.

Ultimately, as thinner UTG will open a wider range of technological applications to electronics, first, a site-specific chemical information must be provided to provide a better understanding of the material and its mechanical and functional behaviours. Here, we demonstrate the importance of using accurate chemical analysis on a commercial glass before and after the strengthening by ion exchange. As shown in the summarized schematic diagram (Fig. 10c), $Li^+$ and $Na^+$ diffuse out from the glass to the $KNO_3$ bath where $K^+$ and impurity $Ca^{2+}$ diffuse into the surface. In comparison with the commonly used chemical analysis instrument, no characteristic peaks of important light and trace amount elements are detected, which this could provide mis-information that may result wrong interpretation in composition-mechanical property relationships.

**Conclusion**

We reported here a comparative study of spectroscopy and atom probe tomography in atomic-scale mapping of alkalis ions in non- and ion-exchanged glass. It shows that APT is superior than commonly used spectroscopy techniques in both spatial resolution and quantification. In addition, direct-flight and reflectron-flight atom probe systems are compared and no significant difference is observed in terms of compositions. In UTG technology, it is very important to quantify and image the exchanged alkali ions in order to increase hardness and flexibility at the same time; therefore, the right mass technique is definitely needed to understand the direct relationship between chemistry and mechanical property.

**Acknowledgement**

B.G., L.T.S. and S.H.K. acknowledge financial support from the ERC-CoG-SHINE-771602. We appreciate Heidi Bögershausen, Petra Ebbinghaus, and Daniel Kurz for Vickers hardness, Raman spectroscopy, ICP-OES measurements. We gratefully acknowledge Jaewon Jang for allowing us to use samples for chemical analysis. Also we thank Uwe Tezins, Christian Broß, and Andreas Sturm for their support to the FIB & APT facilities at MPIE. Dr. Dong-Hyun Lee is gratefully acknowledged for discussion.